\documentclass[a4paper,fleqn,usenatbib]{mnras}

\usepackage{newtxtext,newtxmath}

\usepackage[T1]{fontenc}
\usepackage{ae,aecompl}

\usepackage{graphicx}	
\usepackage{amsmath}	
\usepackage{amssymb}	

\newcommand{\HST}{{\it HST}}

\title[Revisiting the potassium feature in WASP-31b]
{Revisiting the potassium feature of WASP-31b at high-resolution}

\author[N. P. Gibson et al.]{
Neale P. Gibson$^{1}$\thanks{E-mail: n.gibson@qub.ac.uk},
Ernst J. W. de Mooij$^{2}$,
Thomas M. Evans$^{3,4}$,
Stephanie Merritt$^{1}$,\newauthor
Nikolay Nikolov$^{4}$,
David K. Sing$^{5,4}$,
Chris Watson$^{1}$,
\smallskip
\\
$^{1}$Astrophysics Research Centre, School of Mathematics and Physics, Queens University Belfast, Belfast BT7 1NN, UK\\
$^{2}$School of Physical Sciences, Dublin City University, Glasnevin, Dublin 9, Ireland\\
$^{3}$Kavli Institute for Astrophysics and  Space Research, Massachusetts  Institute  of  Technology, Cambridge, MA 02139, USA\\
$^{4}$Physics and Astronomy, University of Exeter, Exeter EX4 4QL,  UK\\
$^{5}$Department of Earth and Planetary Sciences, Johns Hopkins University, Baltimore, MD, USA\\
}

\date{Accepted XXX. Received YYY; in original form ZZZ}

\pubyear{2015}

\begin{document}
\label{firstpage}
\pagerange{\pageref{firstpage}--\pageref{lastpage}}
\maketitle

\begin{abstract}
The analysis and interpretation of exoplanet spectra from time-series observations remains a significant challenge to our current understanding of exoplanet atmospheres, due to the complexities in understanding instrumental systematics. Previous observations of the hot Jupiter WASP-31b using transmission spectroscopy at low-resolution have presented conflicting results.  {\it Hubble Space Telescope (HST)} observations detected a strong potassium feature at high significance ($4.2\sigma$), which subsequent ground-based spectro-photometry with the Very Large Telescope (VLT) failed to reproduce. 
Here, we present high-resolution observations (R>80,000) of WASP-31b with the UVES spectrograph, in an effort to resolve this discrepancy.
We perform a comprehensive search for potassium using differential transit light curves, and integration over the planet's radial velocity.
Our observations do not detect K absorption at the level previously reported with {\it HST}, consistent with the VLT observations.
We measure a differential light curve depth $\Delta F = 0.00031 \pm 0.00036$ using 40\,\AA\ bins centred on the planet's K feature, and set an upper limit on the core line depth of $\Delta F \leq 0.007$ ($3\sigma$) at a few times the resolution limit ($\approx$\,0.24\,\AA).
These results demonstrate that there are still significant limitations to our understanding of instrumental systematics even with our most stable space-based instrumentation, and that care must be taken when extracting narrow band signatures from low-resolution data. Confirming exoplanet features using alternative instruments and methodologies should be a priority, and confronting the limitations of systematics is essential to our future understanding of exoplanet atmospheres.
\end{abstract}




\section{Introduction}

Transmission and emission spectroscopy are powerful techniques to probe the composition and physical properties of transiting planets. Rather than spatially resolving light from the planet, these techniques {\it temporally} separate light from the star and planet. Transmission spectroscopy aims to detect light filtering through the exoplanet's atmosphere during transit, and due to the opacity variations in the atmosphere the planet appears to vary in size with wavelength, allowing recovery of the planet's composition and physical structure \citep{Seager_2000,Brown_2001,Charbonneau_2002}.  Emission spectroscopy isolates the planet's emission by comparing the combined star and planet flux to that measured during secondary eclipse (star only), and can measure the planet's dayside flux as well as phase-resolved emission \citep[e.g.][]{Charbonneau_2005,Deming_2005,Knutson_2007,Cowan_2008}.

These techniques have enabled enormous progress in exoplanet characterisation in recent years, confirming the presence of multiple atomic and molecular features as well as aerosols \citep[see e.g.][]{Crossfield_2015}, and progressing to statistical surveys of hot Jupiters \citep{Sing_2016}.
However, performing these observations has not been straightforward.  At low-resolution, these techniques require ultra-stable time-series to recover flux measurements with precision of $\approx$$10^{-4}$ or better over timescales of $\approx$1\,hour or longer in order to enable useful measurements of exoplanet spectra. However, to date no spectroscopic instruments have been designed to perform such stable time-series measurements, and consequently instrumental systematics limit the precision to typically $\approx$\,$10^{-3}$, leaving our measurements (and therefore our understanding) of exoplanet atmospheres dependent on our ability to model each instrument's unique quirks. Consequently, there is a history of claims and counterclaims in the field \citep[e.g.][]{Sing_2009,Gibson_2011,Deming_2013}.

Despite the increasing use of more sophisticated statistical techniques to account for instrumental systematics \citep{Carter_2009,Gibson_2012,Waldmann_2012,Gibson_2014}, these problems remain. Inference of exoplanet spectra remains largely dependent on the algorithms and modelling inputs chosen by the observers to account for the systematics, and uncertainties in the final spectra rarely propagate all sources of uncertainty, or include estimates of covariance between neighbouring data points. Therefore, despite obvious progress made in recent years, detection and interpretation of spectroscopic features in exoplanet atmospheres is often subjective and varies between different groups.

Observations at high spectral resolution offer an alternative to ultra-stable time-series observations, by using the Doppler shift resulting from the planet's high velocity to disentangle its spectrum from signals that are (quasi-)fixed in wavelength such as stellar and telluric lines \citep{Snellen_2010,Brogi_2012,Birkby_2013}. This technique is particularly effective for detecting molecular features, where resolving many individual lines enables cross-correlation techniques to boost signals.
High-resolution observations are also capable of resolving individual, strong, broad lines, such as sodium and potassium through differential transits \citep[e.g.][]{Snellen_2008,Redfield_2008}, and the technique is rapidly growing in popularity \citep[e.g.][]{Cauley2016,Casasayas-Barris2017,Wyttenbach2017,Khalafinejad_2017,Yan2018}. While insensitive to continuum features in the planet, and necessarily sacrificing signal-to-noise to observe at high resolution, this technique is arguably less sensitive to instrumental systematics, making it well suited to confirming features discovered at low-resolution. For these reasons, we have begun a campaign to re-observe exoplanet transmission spectra where strong features have been detected at low-resolution, to both confirm the signal and to provide resolved measurements to complement the low-resolution spectra.

Here, we present results of observations of the exoplanet WASP-31b at high-resolution, using the UV-Visual Echelle Spectrograph (UVES) at the Very Large Telescope (VLT) \citep{UVES}, which has been previously shown to be sensitive to signatures of exoplanet atmospheres  \citep[e.g.][]{Snellen2004,Czesla2015,Khalafinejad_2017}. WASP-31b is an inflated hot-Jupiter discovered by \citet{Anderson_2011}, with a mass and radius of $\approx0.48\,M_{\rm J}$ and $1.55\,R_{\rm J}$, respectively. It orbits a late F-type dwarf ($V=11.7$) with a period of 3.4\,days.
The optical transmission spectrum of WASP-31b was previously observed at low-resolution using both the Space Telescope Imaging Spectrograph (STIS) on the {\it Hubble Space Telescope (\HST)} and the FOcal Reducer and low dispersion Spectrograph 2 (FORS2) on the VLT. Both sets of observations recovered consistent values for the continuum absorption, revealing evidence for aerosols in the form of clouds (grey scattering) and hazes (Rayleigh scattering) \citep{Sing_2015, Gibson_2017}. However, the STIS observations revealed a significant detection of potassium above the cloud deck (at 4.2\,$\sigma$), whereas the FORS2 observations ruled this out at high-significance. On the one hand \HST/STIS has proven to be one of the most stable instruments for optical transmission spectroscopy, to date performing the largest survey of hot Jupiter atmospheres \citep{Sing_2016}. In addition, the Na feature of WASP-39b detected with \HST/STIS has been confirmed using ground-based FORS2 observations \citep{Nikolov_2016}, and several other features detected with STIS have been later confirmed \citep[e.g.][]{Charbonneau_2002,Snellen_2008,Sing_2008}, including the broadband features of WASP-31b \citep{Gibson_2017}. On the other hand it is extremely difficult to construct an explanation of how the large K signal could be hidden in the FORS2 observations. The importance of \HST/STIS to our current understanding of exoplanet atmospheres makes it even more important that we solve this discrepancy.

One possibility is that unresolved telluric features conspire to hide strong potassium cores where most of the signal could be concentrated \citep{Gibson_2017}. In this case the signal would be considerably larger than measured with \HST, and would therefore be easy to measure at high-resolution. Confirming the potassium signal would also enable us to resolve the feature and determine the pressure structure in the atmosphere. Conversely, failing to recover the signal would have implications for our current observations of exoplanet atmospheres. WASP-31b was therefore selected as our first target for our UVES program.

This paper is structured as follows. In Sect.~\ref{observations} we describe the observations and data reduction, in Sect.~\ref{Analysis} we describe the search for narrow band features, and we discuss the implications of our results in Sects.~\ref{Discussion} and \ref{Conclusions}.

\section{UVES Observations}
\label{observations}

Two transits of WASP-31b were observed using the 8.2-m `Kueyen'  telescope (Unit Telescope 2) of the VLT with UVES: a high-resolution echelle spectrograph \citep{UVES}. Observations were taken on the nights of 2017 April 8 and 2017 May 19, as part of program 099.C-0763 (PI: Gibson). Both transits were observed using a `free template' with dichroic \#2 and cross-dispersers \#2 and \#4.  We used central wavelengths of 437\,nm and 680\,nm in the blue and red arms, respectively, and a slit width of 0.4$^{\prime\prime}$ in both arms, giving R$\sim$80,000. The decker height (slit length) was set as 10 and 7.6 arcseconds for the blue and red arm, respectively, to maximise the background region while avoiding overlap in all relevant orders.

Both sets of observations used an exposure time of 200 seconds, and coupled with a readout time of $\approx$25 seconds resulted in a cadence of 225 seconds, with the exception of a few pauses in the observations during the first night (due to a small earthquake). The first transit observations lasted $\approx$5.2\,hours, resulting in 82 exposures, with approximately the first 18 before ingress, 40 in-transit, and 24 after egress. The second transit observations lasted $\approx$4.4\,hours, resulting in 72 exposures, with 17, 42 and 13 before ingress, during transit, and after egress, respectively. The small difference in in-transit exposures was a result in the short pause in observations during the first night.

Data were analysed using a custom pipeline written in {\sc Python}, which performed basic calibrations, and extracted time-series spectra for each spectral order. We first corrected for the `column swap' which affects the longer wavelength red chip in 625kHz readout mode, where two columns of pixels at the centre of the CCD are mistakenly stored at the edges of the image. Data were also processed using the ESO/UVES pipeline ({version 5.7.0}), which we used for wavelength calibration, and a cross-check of our pipeline. The main advantage of our custom pipeline was that we had a lot more control over the extraction windows and centring process (in particular for optimal extraction) and in the final data products produced (e.g. extracting raw, individual orders using the UVES pipeline required `debug' mode). This proved particularly important when comparing data between orders. We tested various extraction methods to create stable time-series, and in the end used a simple aperture extraction (of width 12 pixels) after sky subtraction. However, we first performed cosmic ray rejection using an optimal extraction algorithm to construct the virtual PSF and replace outlying pixels.

\begin{figure*}
\centering
\includegraphics[width=160mm]{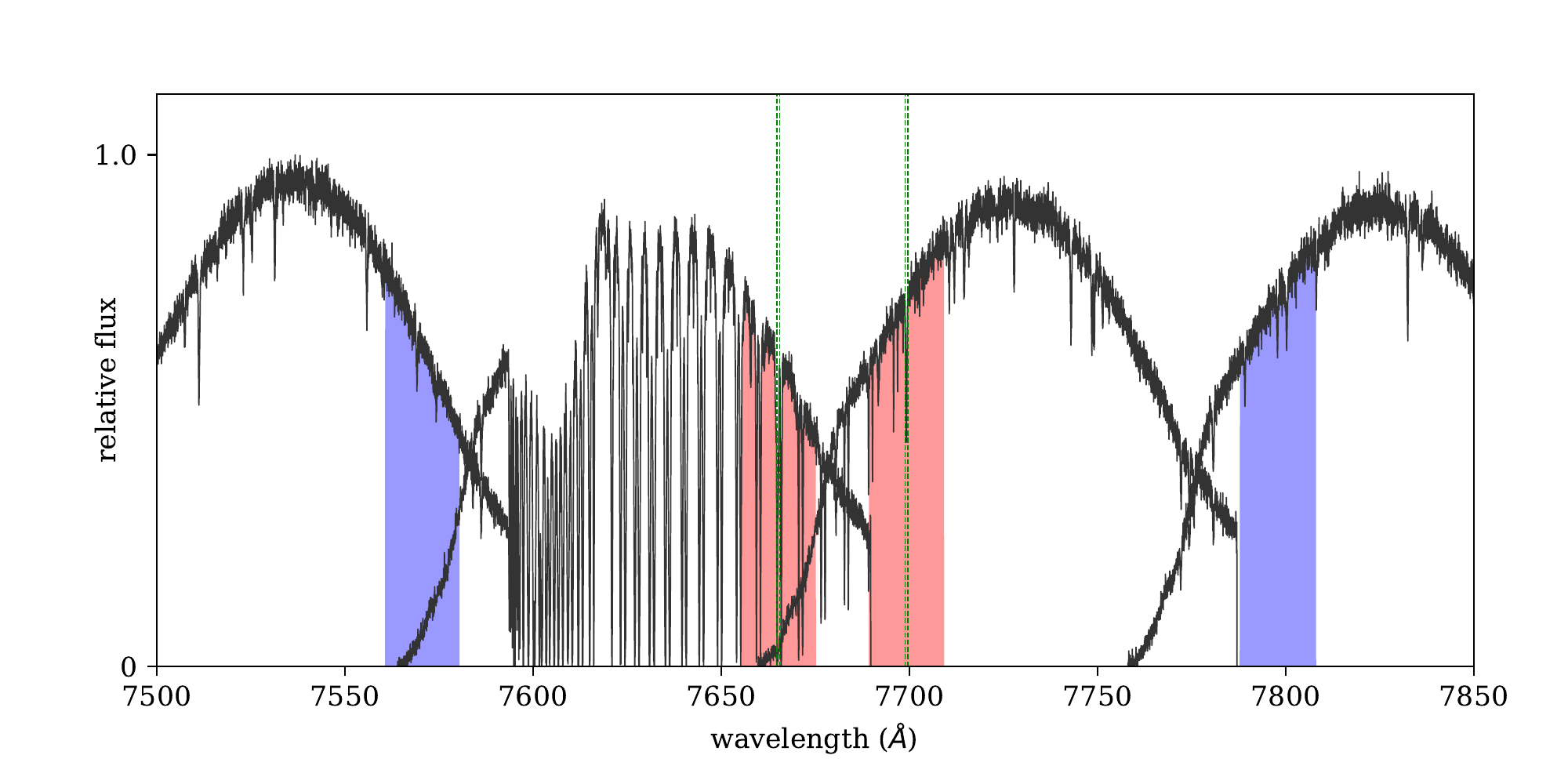}
\caption{Raw spectra of four orders near the potassium doublet. The dashed green lines show the position of the line centres at ingress and egress for absorption at the planet's velocity, after correction for the barycentric and systemic velocity. The red and blue shading show the windows used for constructing the differential light curves (here the bin width is 20\,\AA), with the red windows containing a K line and the blue windows being the corresponding region on the neighbouring order.}
\label{fig:spectra}
\end{figure*}

We proceeded to use the raw flux from the extracted orders, rather than merge orders together via resampling combined with flat-fielding and blaze correction, as the response changes as a function of time and cannot be corrected via flat-fielding. We also checked the stability of the wavelength solution by cross-correlation of the spectral time-series using deep telluric and stellar lines, finding the total drift was less than half a pixel over each night. Fig.~\ref{fig:spectra} shows example spectra of four spectral orders covering the K doublet that were used in the analysis. The mean S/N per pixel near the K feature (taken at the centre of the orders) is $\approx$50 and $\approx$60 for the first and second nights, respectively. Finally, we computed the Barycentric Julian Date for each observation\footnote{Using {\sc Astropy.time} routines.}.

\subsection{Differential light curves}

Our high resolution observations enable us to resolve the individual potassium lines in the spectra, whereas previous low-resolution observations span the full spectral region containing the doublet (typically $\gtrsim$\,40\,\AA). We first proceeded by constructing differential light curves around the individual lines of the potassium doublet (at 7665 and 7699\AA), using a similar technique to \citet{Snellen_2008}. The bins were centred on each individual line, after adjusting for the system velocity and the (mean) barycentric velocity during transit. We used a series of bin widths ranging from 5 to 40\,\AA.
Variations in the barycentric correction and planet velocity during transit are small compared to the bin widths, and were ignored so that the same pixels were used to construct the light curves.
An example of the extraction regions for each order are shown in Fig.\ref{fig:spectra}, and the vertical lines show the extent of the wavelength shifts due to planet's changing velocity during the transit, taking into account the barycentric velocity and the systemic velocity.

To correct for the changing throughput of the Earth's atmosphere and the instrument, we divided these raw light curves by reference light curves taken from different regions of the spectra. This will remove the overall transit shape of the light curve, but will reveal any changes in the transit depth caused by excess absorption in the K bins. We experimented with multiple techniques to achieve the best correction.

We first tested using nearby regions on the spectrum, with bins either side of the each K line. However, we first needed to correct for any blaze variations that can distort the shape of the spectrum as a function of time \citep[e.g.][]{Snellen_2008}. This was achieved by dividing through each spectrum by a master spectrum constructed using a median of the time-series. This removed the dominant spectral features and smoothed each spectra. We then fitted a high-order polynomial (as a function of pixel-value, testing orders varying from 5-15) to each after masking any strong stellar and telluric lines which are not perfectly removed by division of the master spectrum. Finally, each (raw) spectrum was divided through by the fitted polynomial to correct for blaze distortions. Note that this did not remove the blaze function, but rather placed the full time series for each order on a `common' blaze. We then constructed differential light curves from the K bins and neighbouring wavelength channels (of the same width), which provided an adequate correction to probe for the K feature.

\begin{figure}
\centering
\includegraphics[width=90mm]{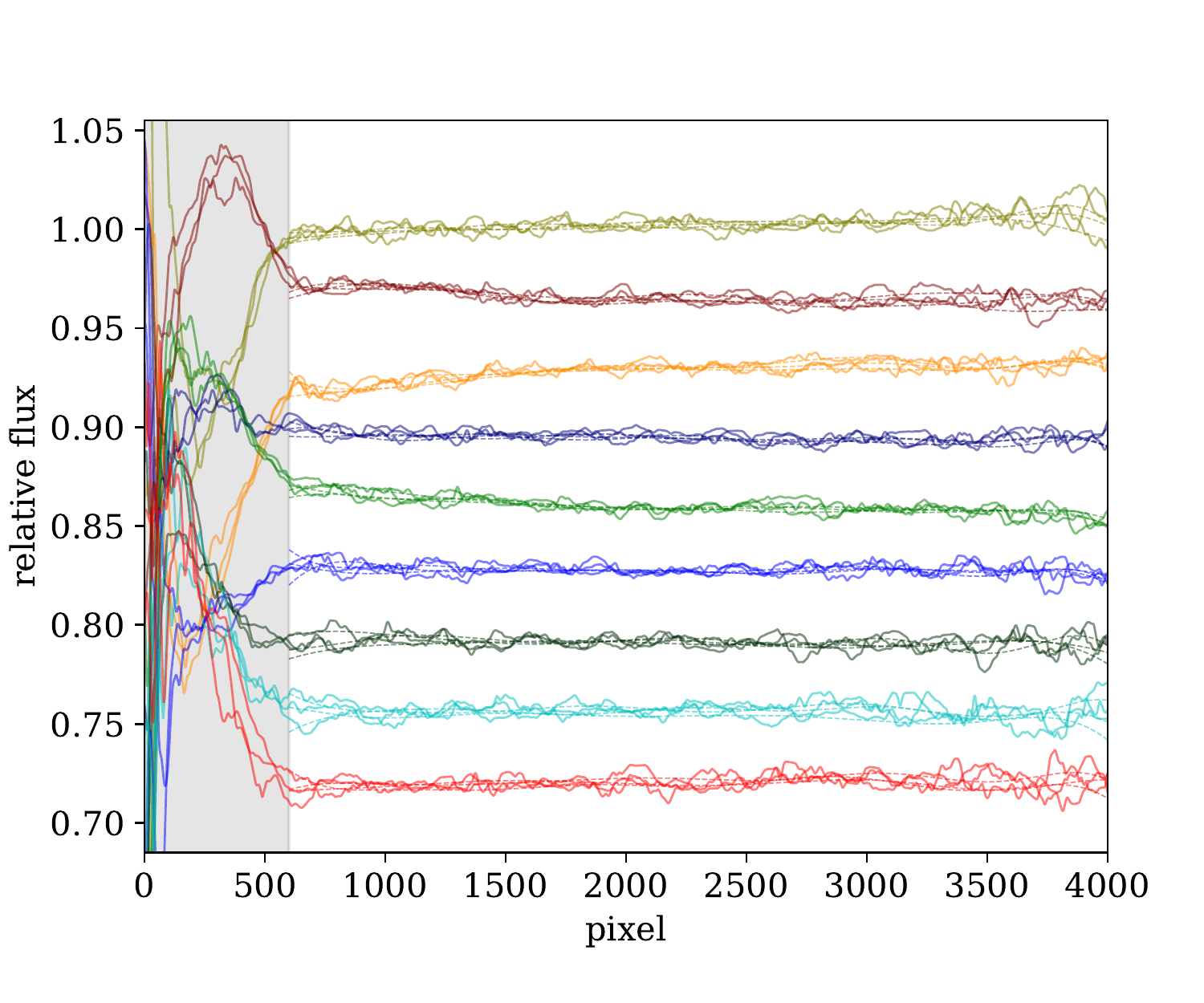}
\caption{A selection of spectra after dividing through by their respective master spectrum (see text), demonstrating the temporal variations but order-to-order stability of the blaze function. The spectra are ordered in time from top to bottom (spaced throughout the first transit observation), and plotted as a function of pixel number on the detector, with the dotted lines showing an 11th order polynomial fit to each. For each spectrum, we show three orders around and including the 7699\,\AA\ potassium line, and normalise each order to that one. The spectra are first median and then Gaussian filtered to reveal the structure in the differential blaze, showing that the neighbouring orders have similar behaviour for each exposure, but change substantially in time. In particular, the blaze function is particularly unstable below $\approx$600 pixels, marked by the shaded region and excluding from the polynomial fit. We excluded the order containing the 7665\,\AA\ potassium line, as the differential spectra are highly variable due to the saturated telluric features.}
\label{fig:blaze}
\end{figure}

The difficulty with this approach is that it limits the size of the bin width that can be used for each extraction, as it is possible to filter out the broad K feature if too high an order correction is used, which is difficult to ascertain.
We instead opted to use a slightly different approach that did not rely on explicit blaze corrections. While the blaze does vary with time, we noted that it does not vary (strongly) with order, and therefore follows a similar temporal variation for neighbouring orders.
Fig.~\ref{fig:blaze} shows a selection of (smoothed) spectra divided through by its corresponding master spectrum for a range of times and spectral orders, demonstrating both the temporal variation and the order-to-order stability of the blaze function.
We therefore opted to use the equivalent pixels on neighbouring orders to construct the reference light curves, and divided through the K light curves by them to correct for atmospheric throughput variations. This provided an extremely simple and robust method to correct the light curves, and requires very few choices to be made in the reduction procedure which could bias the results. Furthermore, this technique uses nearby spectral regions, which minimises systematic effects due to wavelength dependent variations in the Earth's atmospheric throughput and stellar PSF. Finally, using minimum bin widths of 5\,\AA\ limits the impact of line shape variations on the light curves (due to e.g. Centre-to-Limb variations or the Rossiter-McLaughlin effect, \citealt{Snellen_2008}; \citealt{Czesla2015}).

\begin{figure}
\centering
\includegraphics[width=90mm]{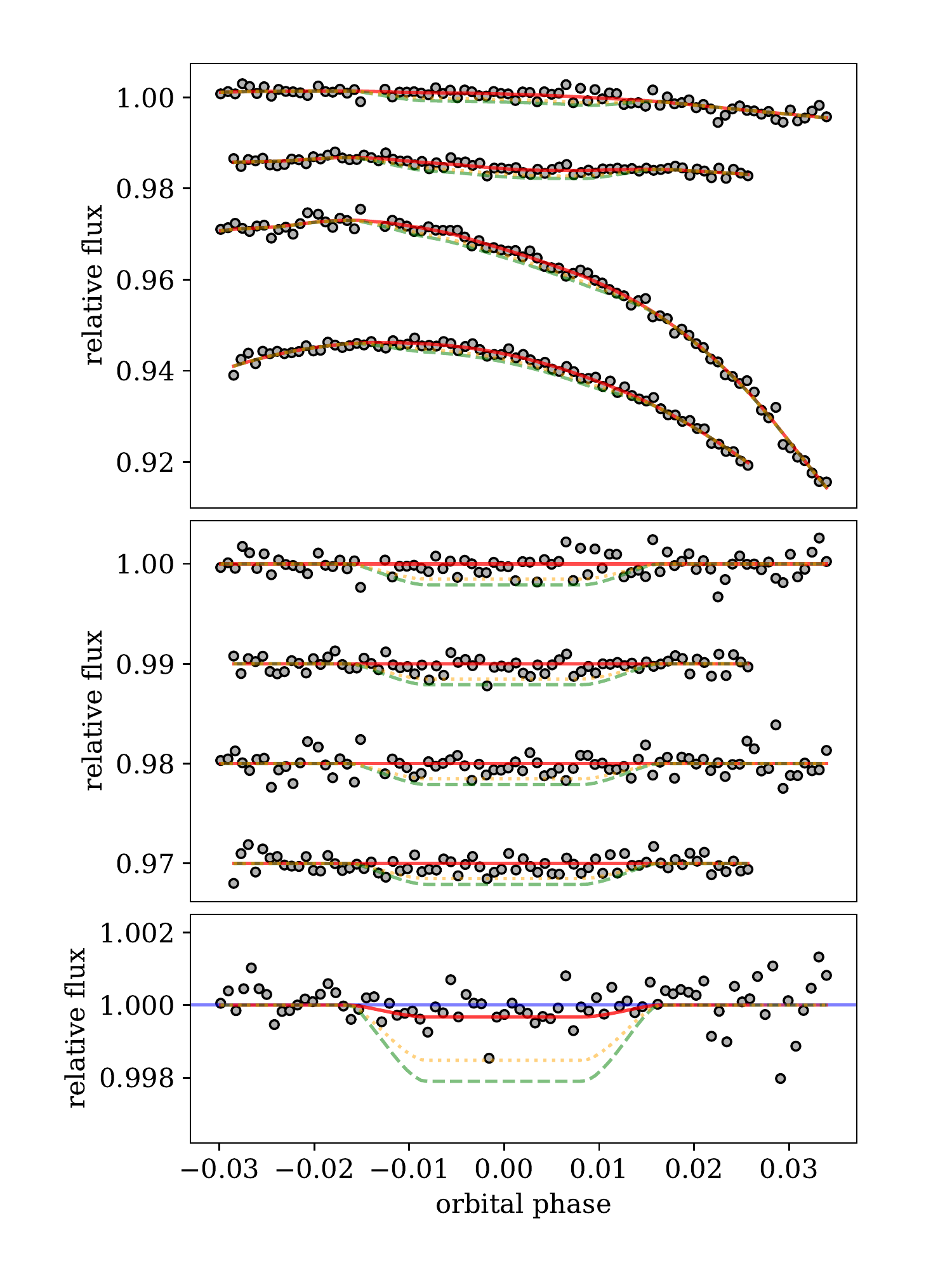}
\caption{Differential light curves of WASP-31b centred around the potassium lines using a bin width of 40\,\AA. Top panel: raw light curves. The top two light curves show the 7699\,\AA\ potassium line for the first and second transit, respectively, and the bottom two are for the 7665\,\AA\ potassium line, which show much stronger airmass trends due to nearby (but not overlapping) telluric lines. The red line shows the best-fit models (see text), the blue shows a null eclipse, and the green and orange lines show the projected detection with STIS using the linear-basis and GP models, respectively. Middle panel: same but with the airmass trends removed. Bottom panel: Phase folded light curve after removing airmass and systematic trends.}
\label{fig:K_lightcurves}
\end{figure}

An example of the comparison bins are also marked in Fig.\ref{fig:spectra}. Note that we only used one adjacent order to avoid potential overlap with the other K line. The regions used in the neighbouring orders are relatively clear of stellar and telluric absorption, and therefore this correction provided an adequate correction without depending on parameters of the blaze correction. The resulting light curves for 40\,\AA\ bins are shown in Fig.~\ref{fig:K_lightcurves}, for both K lines and for both transits.

\section{Analysis}
\label{Analysis}
\subsection{Differential Light Curves of the Potassium Feature}

The differential light curves are of high enough quality to detect excess K absorption even by eye at the level detected using STIS, but no obvious signs of the K feature are visible. There are some prominent systematic effects in the light curves. The dominant one is correlated with the airmass, and is particularly strong for the bins centred at the 7665\,\AA\ K line, where there are deep O$_2$ telluric lines near the K feature. This is caused by the significantly different behaviour of the atmospheric throughput within telluric features. Nonetheless, this variation is on different timescales from the transit, and can be easily separated and accounted for. Note that the 7665\,\AA\ K feature is never completely obscured by any of the telluric lines although is located between two deep lines. In addition, during each transit the planet's velocity moves the absorption lines (by $\approx0.75$\,\AA\ near the K feature) more than the Full Width Half Max (FWHM) of any telluric lines ($\lesssim$0.2\,\AA), meaning the K core in the planet cannot be fully obscured for the duration of a transit. The barycentric correction is also substantially different for the two nights ($\approx$0.4\,\AA) and shifts the lines into different regions.
The 7699\,\AA\ potassium does not overlap any significant telluric features.

We proceeded to fit the light curves using a \citet{Mandel_Agol_2002} light curve model with no limb-darkening, using the transit parameters fixed to those provided in \citet{Sing_2015}. This fixed the duration and shape of the differential transit and we simply fitted for the depth using a single parameter ($\Delta F$). We did not account for the effects of differing limb darkening between the reference wavelength channels. We used the ephemeris of \citet{Sing_2015}, which was accurate to $\approx$3.5 minutes (1$\sigma$) at the epoch of our observations. To model the baseline we used a quadratic function of time, which was able to reliably model the strong airmass trend. We also explored using airmass directly, but found no significant differences in our results.

To account for any residual systematics, we used a Gaussian process (GP) as introduced by \citet{Gibson_2012}, to which we refer the reader for details. We used a time-dependent Mat\'ern 3/2 kernel, which is widely used for ground-based observations where time is the sole input to the GP \citep{Gibson_2013a,Gibson_2013b}, and assumes slightly `rougher' systematics than the standard squared-exponential kernel (by imposing a sharper decline in covariance with time; see \citealt{Gibson_2013a} for details). The covariance kernel (i.e. covariance between two points $i$ and $j$) is dependent only on the time $t$ and is given by:
\[
k(t_i,t_j) = {\xi^2} (1+\sqrt{3}\eta\Delta t) \exp(-\sqrt{3}\eta\Delta t) + \delta_{ij}\sigma^2,
\]
where $\Delta t = |t_i - t_j|$, $\xi$ is the height scale, $\eta$ is the inverse length scale, $\sigma$ is the white noise term, and $\delta$ is the Kronecker delta. In practice we fit for the natural log of the height scale and inverse height scale, which is equivalent to placing priors of the form $p(x) = 1/x$, and is a natural parameterisation for strictly nonzero scale factors.

We modelled all four differential transits simultaneously, using a common depth parameter, and a different baseline model and noise model. This resulted in 1 global parameter, plus 3 baseline variables (quadratic in time including offset) and 3 (correlated-)noise parameters per light curve, giving 25 free parameters in total. We first found a global optimum of the posterior distribution using a differential evolution algorithm\footnote{as implemented in the {\sc SciPy} package, based on \citet{Storn_1997}}, and then used a Differential Evolution Markov Chain (DE-MC) algorithm to refine the solution and explore the posterior distribution in order to extract marginalised distributions (and therefore uncertainties) for the differential transit depth \citep{DEMC,Eastman2013}. We used 80 chains of length 100,000, discarded the first 40\% of each chain, and confirmed convergence using the Gelman \& Rubin statistic \citep{GelmanRubin_1992} after splitting the chains into four groups. We repeated this procedure for all of the extracted wavelength bins.

\begin{figure}
\centering
\includegraphics[width=90mm]{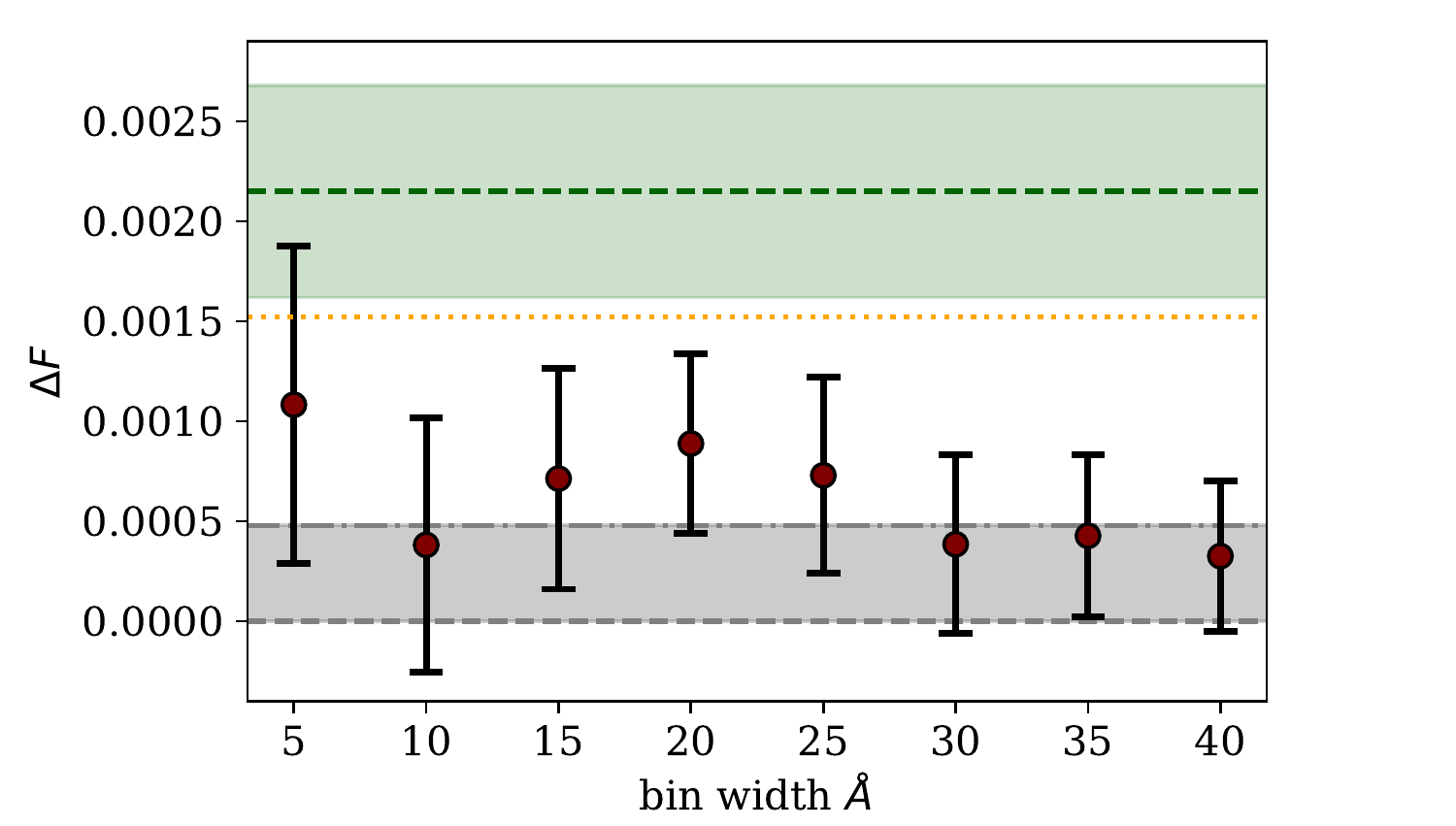}
\caption{Differential light curve depth of the K feature measured for the two lines and both transits as a function of extraction bin width. The horizontal green and orange lines show the detected amplitude of the feature from \HST/STIS using a bin width of 78\,\AA, using the linear-basis and GP models, respectively. The green shading shows the associated uncertainty, and the grey shading marks the 3\,$\sigma$ upper limit set with FORS2.}
\label{fig:k_depths}
\end{figure}

Fig.~\ref{fig:K_lightcurves} shows the best fit models for the raw light curves extracted using 40\,\AA\ bins, as well as after removal of the systematics model (baseline and stochastic GP component). For visualisation, we also phase-folded the light curves. The dashed green and orange lines show the expected feature from the STIS results (for the linear-basis and GP analyses, respectively). The results for the differential transit depth are shown in Fig.~\ref{fig:k_depths}  and Tab.~\ref{tab:results}. The UVES data clearly rule out a K feature at the level inferred from STIS, and we discuss the interpretation of this result in Sect.~\ref{Discussion}.

\begin{table}
\caption{Differential transit depth measured as a function of bin width.}
\label{tab:results}
\begin{tabular}{cc}
\hline
\noalign{\smallskip}
\smallskip
Bin Width (\AA) & $\Delta F$ \\
\hline
5 & 0.00111 $\pm$ 0.00079 \\
10 & 0.00036 $\pm$ 0.00062 \\
15 & 0.00068 $\pm$ 0.00055 \\
20 & 0.00090 $\pm$ 0.00045 \\
25 & 0.00073 $\pm$ 0.00051 \\
30 & 0.00037 $\pm$ 0.00044 \\
35 & 0.00042 $\pm$ 0.00039 \\
40 & 0.00031 $\pm$ 0.00036 \\
\hline
\end{tabular}
\end{table}

\subsection{High-Resolution Search for Individual Potassium Lines}

As well as being able to rule out absorption using relatively wide bins, we can also use our high-resolution data to search for K lines at the resolution limit of the observations. While not quite having sufficient signal-to-noise to detect individual lines of depth $\approx2\times10^{-3}$ in individual pixel channels, the K absorption would be expected to be concentrated within the cores of the lines and therefore much deeper than when diluted over a broad wavelength region of 78\,\AA\ as used in previous measurements. We might expect line depths of >1\% if the lines originated in an extended region of the planet's atmosphere. Furthermore, we can sum over the lines (via cross-correlation) to further increase the signal. We therefore decided to investigate this possibility by looking for evidence of deep and narrow lines in the spectrum.

\begin{figure}
\centering
\includegraphics[width=90mm]{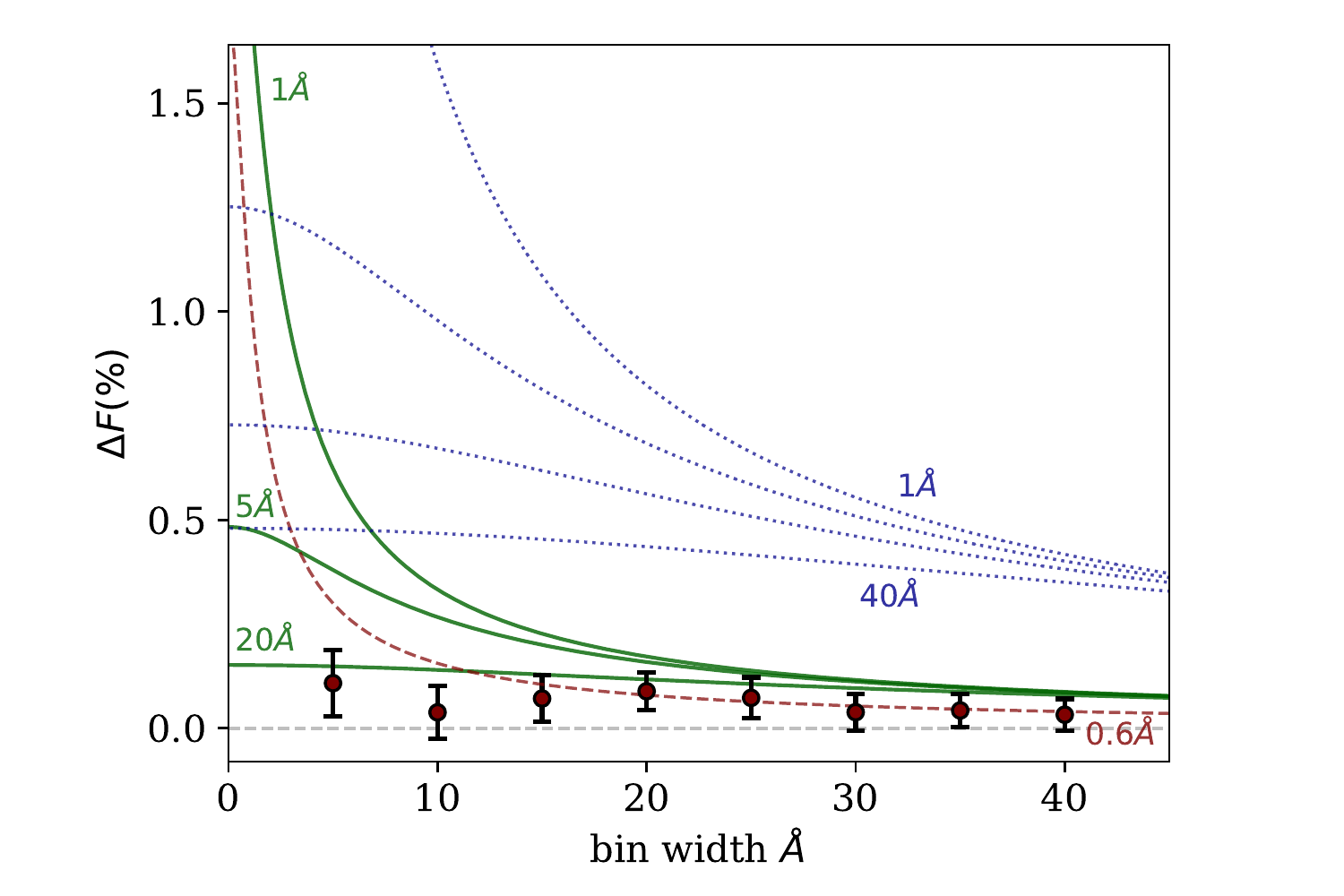}
\caption{Depth of the K feature simulated by integration over a Lorentzian line profile with a cloud deck. The solid lines show the expected growth in differential transit depth for FWHMs of 1, 5 and 20\,\AA, scaled to the 3\,$\sigma$ upper limit set with FORS2 using a bin width of 78\,\AA. The dotted line shows the same for profiles with FWHM of 1, 10, 20 and 40\,\AA, scaled to \HST/STIS measurements using 78\,\AA\ bins (and the linear-basis model analysis), showing that the UVES data are more constraining if the K feature is even moderately narrower than the 78\,\AA\ bin width. Finally, the dashed line shows the simulated K feature for a FWHM of 0.6\,\AA, with arbitrary scaling to show that a strong (>1\%) but very narrow K feature could still be consistent with the UVES differential light curves (red points).}
\label{fig:k_depths2}
\end{figure}

First, in order to illustrate how the signal size might increase for narrower channels, we simulated the K absorption using a Lorentzian line profile, and assumed a flat continuum (e.g. cloud deck) outside the central 78\,\AA\ region (i.e. we set the continuum values to those of the Lorentzian profile at $\pm$38\,\AA). We then integrated over the line profile as a function of bin width (from 0.1--80\,\AA), and measured the signal relative to the continuum. The results are shown in Fig.~\ref{fig:k_depths2} for a range of FWHMs. These simulations make simple assumptions about the line profile and potential cloud deck, but nonetheless clearly demonstrate that for relatively narrow line profiles, the signal is expected to increase rapidly. This means that the UVES results are even less consistent with the STIS measurements, particularly when combined with the detection of aerosols in WASP-31b \citep{Sing_2015, Gibson_2017} that indicate that we would not see strongly pressure broadened wings spanning 10s of \AA. Fig.~\ref{fig:k_depths2} also demonstrates that it is plausible that a strong (>1\%) but narrow (FWHM$\sim$0.6\,\AA) feature could still be consistent with the UVES differential light curves.

We proceeded using methods similar to those used in high-resolution searches for molecular features using cross-correlation techniques \citep{Snellen_2010,Brogi_2012,Birkby_2013}. We started with the raw, extracted orders and implemented a blaze correction as before. However, rather than using the polynomial to fit the differential spectra, we used a median filter followed by a gaussian smoothing with a width of 150 pixels. This provided similar results, but more aggressively removes broad features in the spectra which is acceptable when searching for narrow features. The blaze correction is poorly behaved at the short wavelength end of the spectra (Fig.~\ref{fig:blaze}), and we therefore excluded the first 600 pixels of each order from subsequent analysis. We then applied the {\sc SysRem} algorithm \citep{sysrem} to remove the stellar and telluric features that are essentially fixed in wavelength. This has origins in ground-based photometric surveys but has recently been applied successfully for similar high-resolution time-series \citep{Birkby_2013,Birkby_2017,Nugroho_2017}. We implemented this by treating each wavelength channel as a light curve with uncertainties determined from the Poisson noise of the integrated counts, including the readout noise and sky background. For each pass of {\sc SysRem}, a time-varying vector representing a common trend (representing e.g. airmass, atmospheric throughput, although not necessarily a specific physical parameter) and a vector containing a corresponding co-efficient for each light curve is iteratively determined, before removing from the data. We refer the reader to the \citep{Birkby_2017} and references therein for more details.

\begin{figure*}
\centering
\includegraphics[width=185mm]{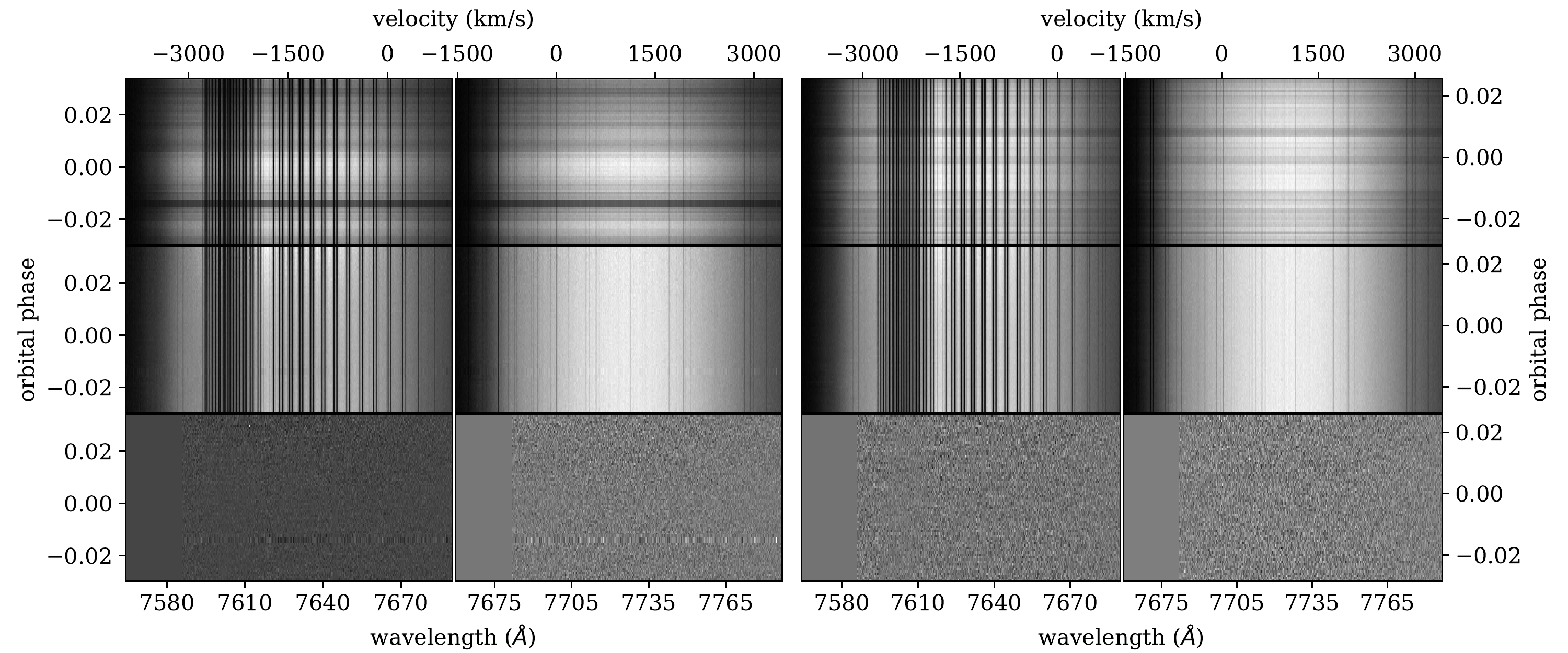}
\caption{Data processing applied to the orders containing the K lines for the high-resolution search. Top panel: raw spectra. Middle panel: after blaze correction. Bottom panel: after application of {\sc SysRem}. This is prior to dividing through by the standard deviation per pixel column. The horizontal artefact in the first transit is the result of the pause in observations due to a small earthquake.}
\label{fig:processing}
\end{figure*}

Only one pass of {\sc SysRem} was required for the order containing the 7699\,\AA\ line, and two passes were required for the other, where stronger telluric features are present. Further passes of {\sc SysRem} did not significantly change our results.
Aggressive removal of stellar and telluric features could in principle begin to remove the planet's signal; indeed, we expect to lose the continuum and any broad features when using this technique (i.e. when a feature doesn't move fast enough through the data as a function of time when compared with it's width), and we can consider (almost any) stellar and telluric removal as a high-pass filter.
However, excessive passes of {\sc SysRem} could even start to remove narrow features \citep[e.g.][]{Birkby_2017}. We checked for this in two ways. First, it was clear by inspecting each step that the dominant signals removed were related to telluric features which are unlikely to correlate with the planet's signature, and therefore narrow planetary features will not be removed. Second, injection tests (described later), gave us further confidence that we were not removing the planet's signal. We note that several alternative methods were tested for removal of the stellar and telluric features, including Principle/Independent Component Analysis and Non-negative Matrix Factorisation, but we found our results insensitive to the exact processing technique used. This demonstrates the stability of UVES for high-resolution time-series observations of exoplanets. Fig.~\ref{fig:processing} shows these steps as applied to the orders containing the potassium lines.

\begin{figure}
\centering
\includegraphics[width=90mm]{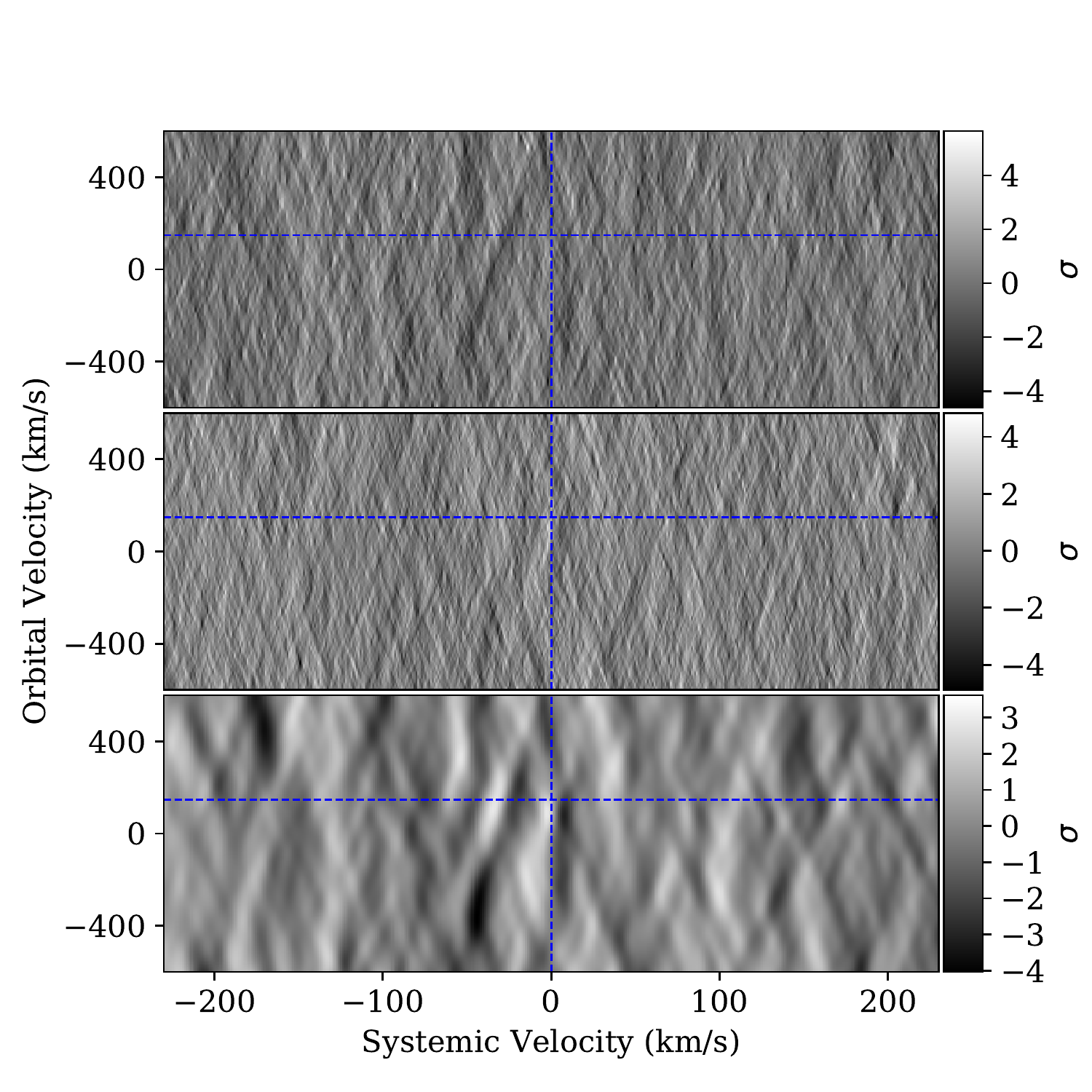}
\caption{Top and middle: Velocity-summed spectra after removing the stellar and telluric signals. Integration was performed for a range of planet semi-amplitudes ($K_{\rm p}$), and the blue lines mark the expected position of the lines. Top and middle panels show the orders containing the 7699\,\AA\ and 7695\,\AA\ lines, respectively. The bottom panel shows the same after cross-correlation with a narrow Gaussian line, and after summing for both lines in the K doublet.}
\label{fig:hr_sig}
\end{figure}

After removing the stellar and telluric lines, we summed up the spectra over the planet's velocity curve, taking into account the time-varying barycentric velocity correction, and weighted using a theoretical transit model (with no limb darkening). To account for varying noise with wavelength, we first divided each column through by its standard deviation. As the planet's exact velocity amplitude ($K_{\rm p}$) depends on the assumed stellar mass, we summed up over a range of velocity amplitudes from $K_{\rm p}$ = -500 to 500 km/s, using a wider range than necessary to assess the impact of any residual systematics and characterise the detection threshold. The expected $K_{\rm p}$ was calculated to be $\approx148\pm9$\,km/s using data from the discovery paper \citep{Anderson_2011}. The resulting cross correlation maps are shown in Fig.~\ref{fig:hr_sig} after summing over both transits for each individual line. No obvious detection of excess absorption is seen near the K lines. These were divided through by their standard deviation after masking the region near the expected K features, which enables an initial, crude estimate of the detection significance.

As any potassium feature will also be present in neighbouring pixels, cross-correlation with a template spectrum should in principle increase the detection significance. We therefore cross-correlated the spectra with a Gaussian of FWHM=0.24\,\AA\ centred at the expected line position in order to integrate the signal over a narrow feature. This corresponds to a few times the resolution element of the instrument. We then summed up the cross-correlation signals over the planet's velocity curve as before. The resulting cross correlation maps are shown in Fig.~\ref{fig:hr_sig} after summing over both transits and for both K lines.  Again, no obvious detection of excess absorption is seen near zero velocity and at the expected $K_{\rm p}$ of the planet. 

\begin{figure}
\centering
\includegraphics[width=90mm]{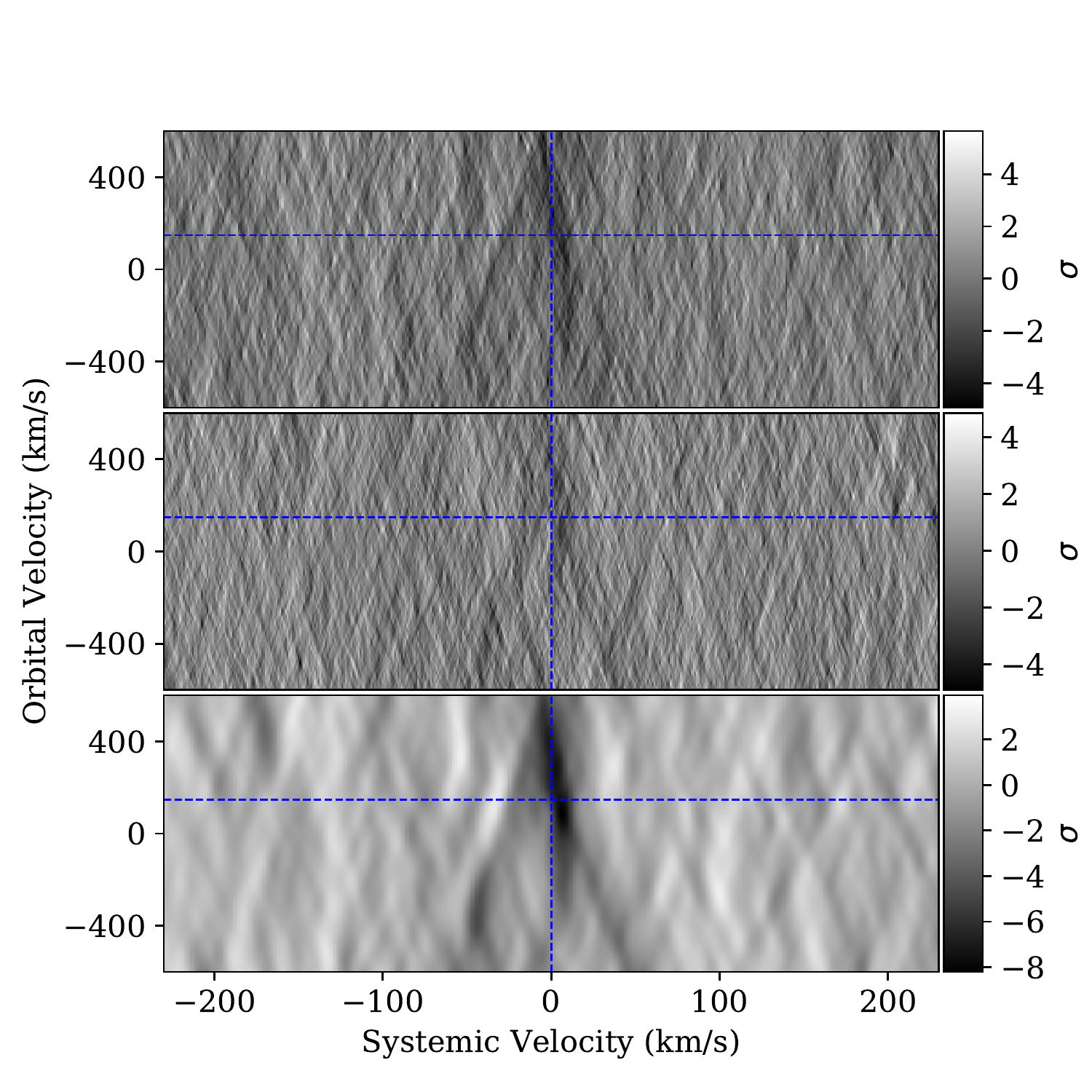}
\caption{Same as Fig.~\ref{fig:hr_sig} with an injected K signal of depth 1\%.}
\label{fig:hr_sig_injected}
\end{figure}

In principle the signal-to-noise is high enough that a K line of $\sim$1\% deep would be clearly detected (depending on the line width). In order to test this we repeated the analysis after injecting a signal into the raw data, i.e. prior to the blaze correction. We injected a line of 1\% depth and FWHM of 0.6\,\AA\, at the expected velocity of the planet ($K_{\rm p}$ = 148 km/s), and processed the data using the same methods as before. The results are shown in Fig.~\ref{fig:hr_sig_injected}. Hints of the absorption lines are visible in the upper two maps (i.e. without cross-correlation), and the signal is clearly detected after cross-correlating using the template lines. This also showed that the {\sc SysRem} algorithm did not remove the planet's signal when removing the stellar and telluric signals.

\begin{figure}
\centering
\includegraphics[width=90mm]{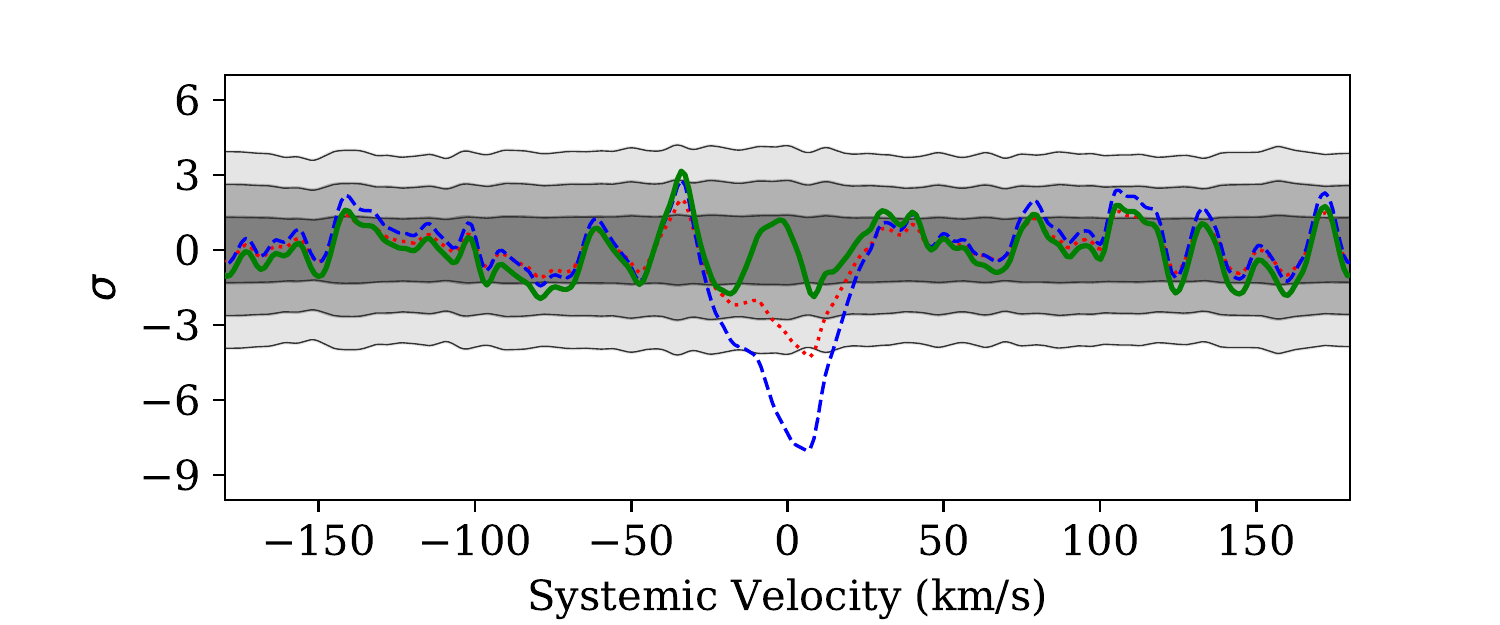}
\caption{Slices through the cross-correlation map at the expected planet velocity. The green line shows the actual signal from the data, and the blue dashed line shows the same after injection of the 1\% K features, clearly indicating that the data can detect such features. The red dotted line shows the approximate $3\,\sigma$ detection threshold with K line depth of 0.7\%.}
\label{fig:slices}
\end{figure}

In order to set a robust detection significance, we repeated the analysis (without injection), after randomising the order of the spectra, following \citet{Esteves_2017}. This was repeated 1000 times and the standard deviation of the cross-correlation map was calculated, and used to set the position dependent detection significance. Fig.~\ref{fig:slices} shows slices of both the real and injected cross-correlation map at the expected $K_{\rm p}$ of the planet. The grey shading marks the 1, 2 and 3\,$\sigma$ confidence intervals determined using the randomised phasing. In order to set an approximate upper limit on the K line depth, we repeated the injection test until we reached a detection above 3\,$\sigma$, which was approximately a line depth of 0.7\%. The exact detection threshold would vary as a function of injected line width. The detection significance would be expected to increase with width unless the core and wings of the line were broad and flat when compared to the planet's velocity shift between exposures, in which case the differential light curve technique would become more appropriate.

\subsection{Search for Na features}

We also performed similar searches for Na features, using both techniques of differential light curves and high-resolution cross-correlation. To extract the Na light curves we used bin widths of 10--40\,\AA. As the lines of the Na doublet are separated by only $\approx$6\,\AA, we used extraction regions centred on mean wavelength of the doublet (after accounting for barycentric velocity and systemic velocity). The light curves were extracted from two overlapping orders, and for both transits, again resulting in 4 light curves per extraction bin.

\begin{figure}
\centering
\includegraphics[width=90mm]{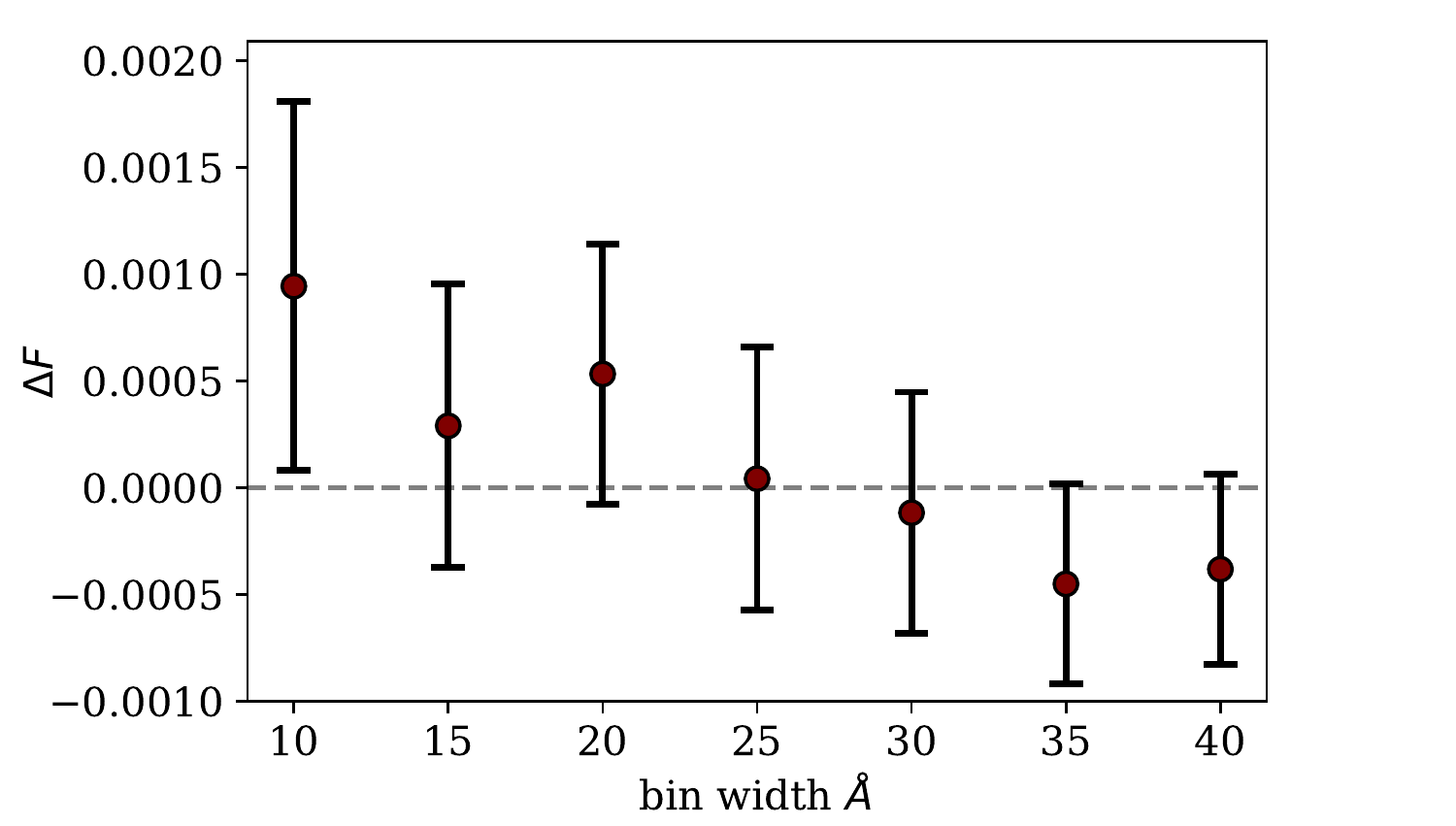}
\caption{Differential light curve depth of the Na feature measured using the combined fit of the doublet and two transits.}
\label{fig:na_depths}
\end{figure}

A similar fitting procedure was applied to fit the light curves simultaneously, and the results for the differential transit depths as a function of bin width is shown in Fig.~\ref{fig:na_depths}. For the widest bin of 40\,\AA, we measured $\Delta F = -0.00038 \pm 0.00045$, showing no evidence of excess Na absorption. We also performed a similar high-resolution search for the Na feature, combining cross-correlation with integration over the radial velocity curve of the planet. Again, we found no evidence for Na absorption.

\section{Discussion}
\label{Discussion}

Transmission spectroscopy of WASP-31b using \HST/STIS revealed the presence of a cloud deck and a Rayleigh scattering feature at short wavelength, and also the presence of a strong potassium feature (4.2\,$\sigma$), but no sodium, which was interpreted as evidence of sub-solar Na/K abundance ratio. Physically, this could be explained by Na condensation depleting it from the upper atmosphere, or alternatively evidence for primordial abundance variations, either in the disk or due to accretion processes \citep{Sing_2015}. Subsequent ground-based observations from VLT/FORS2 confirmed the aerosol features, but failed to reproduce the K feature \citep{Gibson_2017}. Furthermore, a reanalysis of the STIS data with GP methods reduced the detection significance of the K feature to $<3$\,$\sigma$. 
The explanation for this discrepancy was either an underestimate of the uncertainties when using the standard linear-basis model of STIS systematics, or alternatively that the K feature was concentrated in narrow cores which could be (at least in part) hidden by telluric absorption, as the K feature falls near the strong O$_2$ A-band. 

The key result of this paper is that we cannot reproduce the large potassium detection using our high-resolution UVES observations using both differential transit light curves and a cross-correlation method. Our observations measure a differential transit depth of $\Delta F = 0.00031 \pm 0.00036$ using a bin width of 40\,\AA, compared to the STIS value of $\Delta F =  0.00215 \pm 0.00053$ using a bin width of 78\,\AA\ (or $0.0015 \pm 0.0006$ using the GP model). Clearly this rules out the large potassium detection, particularly keeping in mind that it should be even larger within a narrower channel. For example, in Fig.~\ref{fig:k_depths2} we show that even for a relatively broad K feature with FWHM = 40\,\AA\ (with a Lorentzian profile) the differential transit depth at 5\,\AA\ grows to over 0.4\%, and is considerably larger for narrower features. In addition, at high-resolution we can confirm that the K cores do not fall on top of any of the strongest telluric lines, and can rule out tellurics as an explanation of the lack of ground-based detection. Furthermore, the VLT/FORS2 data measured a differential transit depth of $\Delta F = -0.00006\pm0.00018$ using the same bins as the STIS data, placing even more stringent constraints on the potassium feature. Our UVES data indicates that the K cores are both unable to overlap with deep telluric features, confirming that telluric absorption was not responsible for hiding the K feature from the ground-based low-resolution observations with FORS2. Finally, we can rule out very narrow lines of  $>0.7\%$, further ruling out the possibility that the K lines are concentrated in narrow cores that could be hidden by telluric features.

The only remaining explanation for the detection using \HST, is that the systematics model used for STIS is inadequate to describe the data. This has implications for our current understanding of exoplanet atmospheres, as STIS remains the foremost instrument for optical observations of transmission and reflection spectra \citep[e.g.][]{Sing_2016}. However, as systematics in the raw STIS light curves are often at the $10^{-3}$ level or greater (which is typically the case for most time-series spectrocopy), they are challenging to model accurately. This is further complicated by the orbit of \HST, which orbits Earth every $\sim$96 minutes, dividing transit observations in discrete blocks of $\sim$48 minutes. 

STIS time-series systematics are typically fitted using a linear-basis model simultaneously with the transit model. This usually includes up to a 4th order polynomial of phase, a linear function of time, and additional auxiliary variables such as the position of the star over the CCD, or width of the spectral trace \citep{Sing_2015,Nikolov_2015}. For a detailed description of linear basis models as systematics models see \citet{Gibson_2011} and \citet{Gibson_2014}.  There are many potential problems with such models that could lead to underestimated systematic effects. These include using insufficient model complexity, or even inadequate numerical methods to fit the model to the data. A more problematic issue is whether we can model systematics using linear basis models at all, as it is entirely possible that the main drivers of the systematics models are not known or measured, and even when they are, polynomial expansions of these basis inputs may not be sufficient to explain the observations. Indeed, there is precedent for this situation with \HST, where the NICMOS instrument was used for some of the earliest detections of molecular features in exoplanet atmospheres, but the systematics models were subsequently found to be lacking through statistical methods and further observations \citep{Gibson_2011,Gibson_2012,Crouzet_2012,Deming_2013}.

Nonetheless, it is clear that STIS does not have the same scale of issues as NICMOS. Many STIS observations have been confirmed with alternative instrumentation, including the optical transmission spectra of HD 209458b \citep{Charbonneau_2002,Snellen_2008,Sing_2008}, WASP-39b \citep{Nikolov_2016}, and even the broadband features in WASP-31b \citep{Gibson_2017}. It is therefore difficult to diagnose exactly what went wrong with the WASP-31b potassium feature. It is possible that model selection procedures underestimate the complexity of the systematics model required for narrow band channels, where the complexity is typically set via analysis of broad band light curves. This would explain why GP methods did not detect a significant K feature, but were able to recover the same broad-band features with similar uncertainties \citep{Gibson_2017}. GP methods did however still detect an excess of K at $\approx$2.5\,$\sigma$. Although not significant, this might hint that the choice of model inputs is the problem, rather than simply the complexity of the chosen basis model. A final point, is that given our lack of understanding of the true cause of the systematics, it is entirely possible that these simple, ad hoc systematics models will work for some sets of observations, and fail for others, or even fail for specific wavelength channels. This is particularly likely where auxiliary parameters are extrapolated onto the transit, where the linear basis model will be less stable.

It is perhaps unsurprising that given the amplitude of the systematics models compared to our measurement precision, and our inability to understand the root cause of the systematics (at the very least we cannot construct physical models), that from time to time we should expect to get the wrong answer. This does not undermine the case for continuing observations of transiting planets using such methods, but does emphasise the need to perform either repeated transit observations (where systematics are unlikely to be correlated), or, even better, to use alternative ground-based facilities and methodology to confirm detections. Indeed, it is certainly likely that the statistical picture that is emerging of hot Jupiter atmospheres (e.g. dominated by Na, K, water, aerosols) is based in reality; however, individual detections of features and derived quantities based on these (e.g. mixing ratios, metallicities) should be treated with a healthy dose of skepticism and re-observed where possible.

\section{Conclusions}
\label{Conclusions}

We presented new high-resolution observations of two transits of the hot Jupiter WASP-31b with UVES, in an effort to confirm the strong potassium feature previously detected with \HST/STIS, and potentially resolve the shape of the lines. Our observations are precise enough to probe for the K feature, using both differential light curves, and integration over the planet's radial velocity curve after removing stellar and telluric features. This further demonstrates that the UVES spectrograph is well suited for probing atomic and molecular features at high-resolution at optical wavelengths. However, we failed to detect the K feature, and can conclusively rule out the presence of a strong K line in the planet, confirming previous ground-based observations with VLT/FORS2 at low-resolution.

The only remaining explanation for this is that the models used to describe the STIS systematics were inaccurate and/or led to underestimated uncertainties. This raises important questions for our current understanding of exoplanet atmospheres. While previous observations suggest that STIS produces consistent and reliable results, these observations suggest that this is not always the case. 
Such observations highlight the difficultly in modelling instrumental systematics, which is unsurprising considering that they are typically an order-of-magnitude larger than the precision required for useful exoplanet spectroscopy. It is useful to periodically remind ourselves of this sobering reality, and we emphasise the need to obtain repeated observations of exoplanet spectra, using a wide variety of instrumentation and techniques, in order to built up a robust and coherent picture of exoplanet atmospheres.

\section*{Acknowledgements}

This work is based on observations collected at the European Organisation for Astronomical Research in the Southern Hemisphere under ESO programme 096.C-0765. N. P. G. gratefully acknowledges support from the Royal Society in the form of a University Research Fellowship. C. A. W. acknowledges support from STFC grant ST/L000709/1. We are grateful to the developers of the {\sc NumPy, SciPy, Matplotlib, iPython} and {\sc Astropy} packages, which were used extensively in this work \citep{Jones_2001,Hunter_2007,Perez_2007,Astropy}. Finally, we thank the referee for their careful reading of the manuscript and useful suggestions.

%




\bibliography{../MyBibliography} 
\bibliographystyle{mnras} 



%
%


\bsp	
\label{lastpage}
\end{document}